\documentclass{PoS}

\newcommand{\GeV}{\,\mathrm{GeV}}
\newcommand{\GeVpc}{\,\mathrm{GeV/c}}

\title{Energy dependence of transverse particle production from SPS to RHIC}

\ShortTitle{Energy dependence of transverse particle production from SPS to RHIC}

\author{\speaker{Andr\'as L\'aszl\'o}, for the NA49 Collaboration\\
        RMKI, Budapest\\
        E-mail: \email{laszloa@rmki.kfki.hu}}

\abstract{Transverse momentum spectra up to $4.5\GeVpc$ were measured around 
midrapidity in Pb+Pb collisions at $\sqrt{s_{{}_{NN}}}=17.3\GeV$, for 
$\pi^{\pm}$, $p$, $\bar{p}$ and $K^{\pm}$, by the NA49 experiment. 
The nuclear modification factors $R_{AA}$, $R_{AA/pA}$ 
and $R_{CP}$ were extracted and are compared to RHIC results at 
$\sqrt{s_{{}_{NN}}}=200\,\mathrm{GeV}$. The modification factor $R_{AA}$ shows 
a rapid increase with transverse momentum in the covered region. 
The modification factors $R_{AA/pA}$ and $R_{CP}$ shows a saturation 
well below unity in the $\pi^{\pm}$ channel. The extracted $R_{CP}$ values 
follow the $200\,\mathrm{GeV}$ RHIC results closely in the available transverse 
momentum range for all particle species, except for $\pi^{\pm}$. 
For $\pi^{\pm}$ above $2.5\,\mathrm{GeV/c}$ transverse momentum, 
the measured suppression is smaller than that observed at RHIC. 
The $K^{+}/\pi^{+}$ and 
$K^{-}/\pi^{-}$ ratios were also extracted from the data and are compared to 
the $200\,\mathrm{GeV}$ RHIC results. The values of the former ratio is 
above, whereas the values of the latter ratio is slightly below the 
$200\,\mathrm{GeV}$ RHIC points.}

\FullConference{High-pT Physics at LHC - Tokaj'08 \\
		 16-19 March 2008\\
		 Tokaj, Hungary}

\begin{document}

\section{Introduction}

One of the most interesting features discovered at RHIC 
is the suppression of particle production at high transverse momenta in 
central nucleus-nucleus reactions, relative to peripheral ones as well as 
to p+nucleus and to p+p collisions 
\cite{phenixauaunopid,phenixauaupid,phenixdaunopid,phenixdaupid,starauaupid}. 
This is generally interpreted as a sign of parton energy loss in hot and 
dense strongly interacting matter.

The aim of the presented analysis is to investigate the energy dependence of 
these effects via a systematic study of Pb+Pb reactions at top ion-SPS 
energy, $158A\GeV$ 
($\sqrt{s_{{}_{NN}}}=17.3\GeV$), with the CERN-NA49 detector \cite{nim99}. 
A similar study has been published by the CERN-WA98 collaboration 
for the $\pi^{0}$ channel \cite{wa98,wa98new}. Our analysis 
extends the existing results to all charged particle channels, 
i.e.\ $\pi^{\pm}$, $p$, $\bar{p}$ and $K^{\pm}$.

Invariant yields were extracted as a function of transverse momentum 
$p_{{}_T}$ in the range from $0.3$ to $4.5\GeVpc$ in the 
rapidity interval $-0.3\leq y\leq 0.7$ (midrapidity), at 
different collision centralities \cite{hipt,hipttech}. 
Identification of particle types is crucial, because the particle 
composition of hadron spectra changes rapidly with transverse momentum and 
differs significantly from that observed at RHIC energies.

Using the identified single particle spectra from \cite{hipt} and the 
charged pion spectra from \cite{pp,pC,pC2}, the nuclear modification factors 
were calculated. These are defined as
\[R^{{}^{BC}}_{A_{1}+A_{2}/A_{3}+A_{4}}=\frac{\left<N_{{}_{BC}}(A_{3}+A_{4})\right>}{\left<N_{{}_{BC}}(A_{1}+A_{2})\right>}\cdot\frac{\mathrm{yield}(A_{1}+A_{2})}{\mathrm{yield}(A_{3}+A_{4})},\]
\[R^{W}_{A_{1}+A_{2}/A_{3}+A_{4}}=\frac{\left<N_{W}(A_{3}+A_{4})\right>}{\left<N_{W}(A_{1}+A_{2})\right>}\cdot\frac{\mathrm{yield}(A_{1}+A_{2})}{\mathrm{yield}(A_{3}+A_{4})},\]
where $\left<N_{{}_{BC}}\right>$ and $\left<N_{W}\right>$ are, respectively, 
the average number of: binary collisions and wounded nucleons, calculated for 
the reactions $A_{1}+A_{2}$ and $A_{3}+A_{4}$ \cite{VCALrel,VCALabs}. 
$R_{AA}$, $R_{pA}$ and $R_{CP}$ are used to denote the special cases 
$R_{A+A/p+p}$, $R_{p+A/p+p}$ and $R_{\mathrm{Central}/\mathrm{Peripheral}}$ for $A+A$ 
reactions, whereas $R_{AA/pA}$ abbreviates $R_{A+A/p+A}$. These were 
calculated and compared with the $\sqrt{s_{{}_{NN}}}=200\GeV$ RHIC 
results \cite{phenixauaupid,phenixdaupid}.

Also the $K^{+}/\pi^{+}$ and $K^{-}/\pi^{-}$ ratios were extracted from the data 
and are compared to $\sqrt{s_{{}_{NN}}}=200\GeV$ RHIC results of \cite{phenixauaupid}.

\section{Analysis details}

The centrality of events was determined using the energy of 
projectile spectators deposited in a downstream Veto Calorimeter 
(VCAL).
Careful study of the detector response and Glauber calculations were performed 
in order to obtain $\left<N_{{}_{BC}}\right>$ and $\left<N_{W}\right>$ as a function 
of centrality \cite{hipt,VCALrel,VCALabs}.

Due to the rapid decrease of the $p_{{}_T}$ spectra, special care was 
taken to achieve good signal-to-noise ratio in the high-$p_{{}_T}$ 
region. Possible 
background tracks were rejected with a twofold filtering procedure: 
(1) discontinuous tracks were discarded, and (2) tracks originating from 
the acceptance border were rejected \cite{hipt,hipttech}.

Particles were identified at the spectrum level, using specific ionization 
($\frac{\mathrm{d}E}{\mathrm{d}x}$) fits \cite{hipt,hipttech}.

The resulting particle spectra were corrected for feed-down ($5$ to $30\%$), 
decay loss ($20\%$ to $0\%$), tracking inefficiency (below $10\%$), 
non-target contribution (below $5\%$), and geometric acceptance. 
The fake-rate, momentum smearing, and momentum scale uncertainty 
proved to be negligible. The correction details are discussed in \cite{hipt,hipttech}.

After full correction, the systematic errors are about $2.2\%$ for 
$\pi^{+}$, $\pi^{-}$ and $K^{-}$, $3.7\%$ for $p$, $4.5\%$ for $K^{+}$, 
and $6.5\%$ for $\bar{p}$ \cite{hipt,hipttech}.

\section{Results and discussion}

\begin{figure}[!ht]
\begin{minipage}{70mm}
\begin{center}
\includegraphics{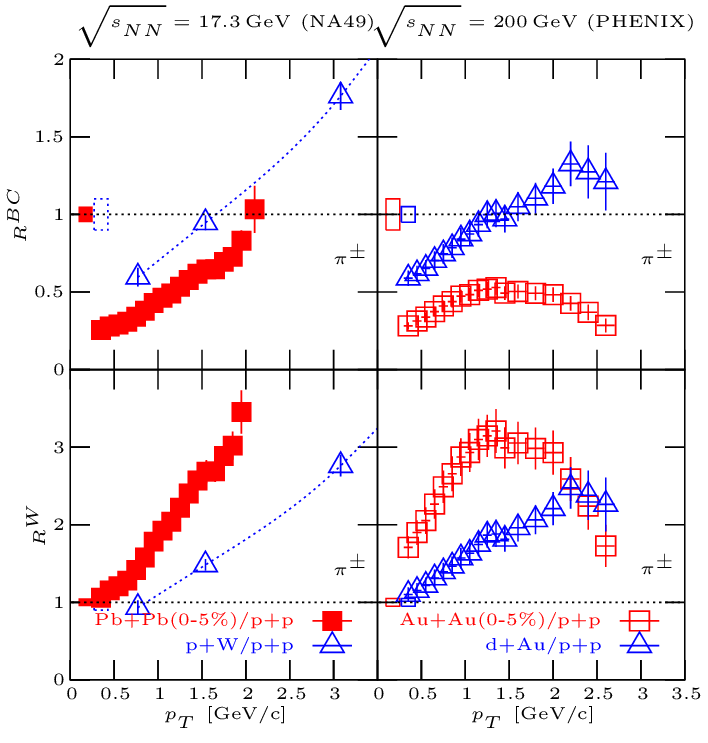}
\includegraphics{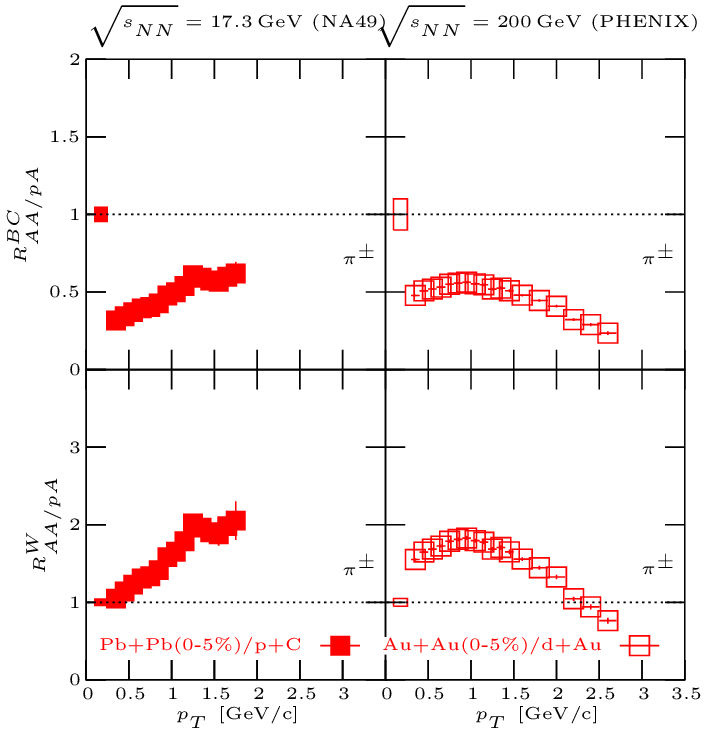}
\end{center}
\end{minipage}\hspace*{2.5mm}%
\begin{minipage}{70mm}
\begin{center}
\includegraphics{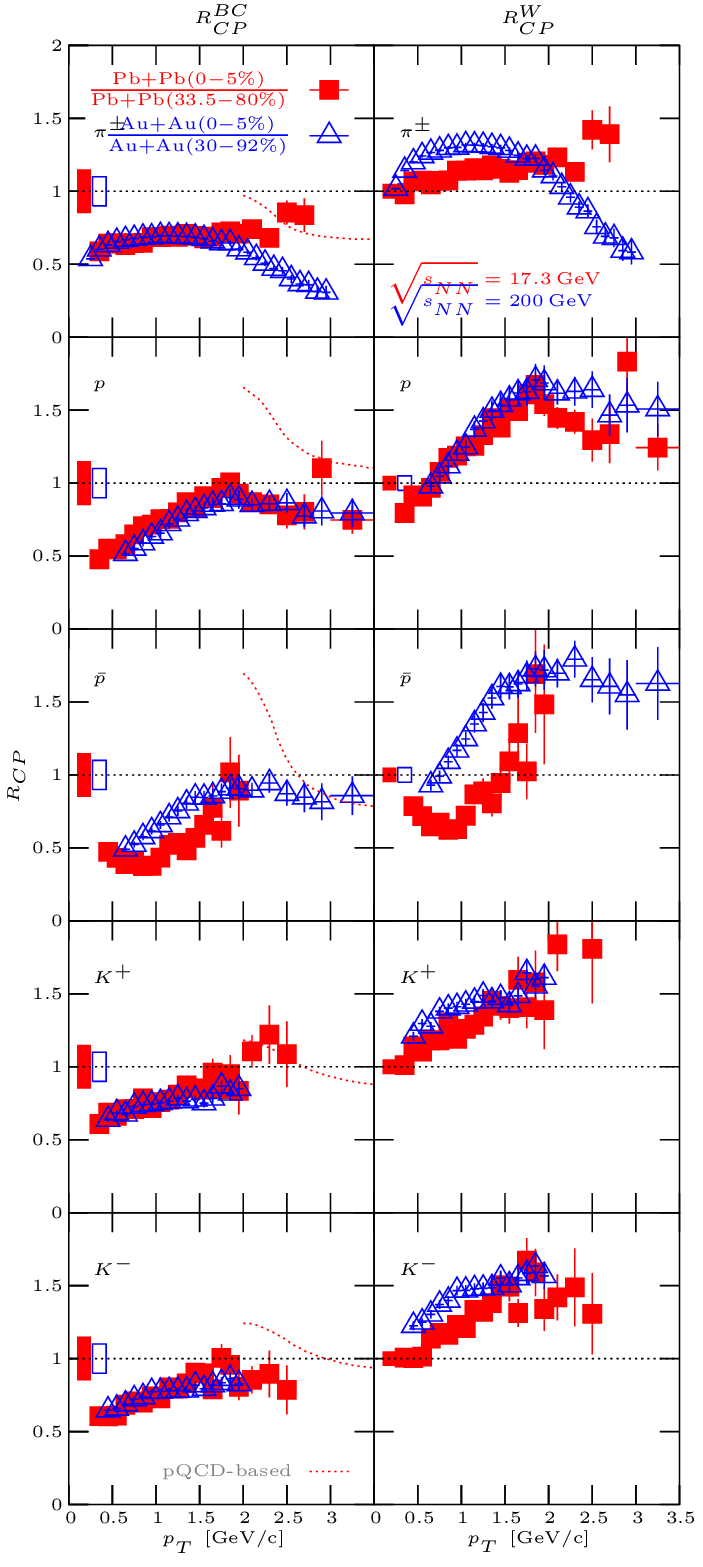}
\end{center}                                                                     
\end{minipage}
\caption{
$R_{{}_{AA}}$ (top left), $R_{{}_{AA/pA}}$ (bottom left), and 
$R_{{}_{CP}}$ (right) nuclear modification factors, measured at 
$\sqrt{s_{{}_{NN}}}=17.3\,\mathrm{GeV}$ \cite{hipt,pp,pC,pC2}, and compared to 
$\sqrt{s_{{}_{NN}}}=200\,\mathrm{GeV}$ data \cite{phenixauaupid,phenixdaupid}. 
To supplement the existing SPS energy data, a $\sqrt{s_{{}_{NN}}}=19.4\,\mathrm{GeV}$ 
p+W/p+p data set was also used \cite{cronin}. The dotted lines in the top left panel 
are drawn to guide the eye. The dotted lines in the right panel 
indicate a pQCD-based energy loss model prediction \cite{xnw}.}
\label{NMF}
\end{figure}

The inclusive particle spectra of \cite{hipt,pp,pC,pC2} allow to calculate the 
nuclear modification factors $R_{AA}$, $R_{pA}$, $R_{AA/pA}$ and $R_{CP}$ for top ion-SPS 
energy. 
Additionally, a close-by energy data set \cite{cronin} was considered
for the p+W/p+p yield ratios. 
These results are compared to similar 
quantities at the top RHIC energy \cite{phenixauaupid,phenixdaupid}. 
All of this is shown in Fig.\ \ref{NMF}, along with pQCD-based energy loss 
model predictions for $R_{CP}$ \cite{xnw}.

It is seen that the $R_{AA}$ and the $R_{pA}$ at the top
ion-SPS energy increase monotonically with $p_{{}_T}$ in the covered
region. This is sometimes referred to as 
the Cronin effect. A widely accepted explanation for this phenomenon is 
initial multiple scattering on either partonic or hadronic level depending 
on the valid particle production picture. It is also observed that the $R_{AA}^{BC}$ 
points stay below $R_{pA}^{BC}$, whereas the $R_{AA}^{W}$ points are above 
$R_{pA}^{W}$ both at top ion-SPS energy and at RHIC. The $R^{W}$ modification 
factors start approximately from unity.

If the multiple scattering interpretation of the Cronin effect is valid, it is natural 
to expect it to contribute much less to the nuclear modification factors $R_{AA/pA}$ 
and $R_{CP}$. Therefore, if looking for 
nuclear effects other than multiple scattering, these ratios should also be considered. 
The $R_{AA/pA}^{BC}$ and $R_{CP}^{BC}$ stay well below unity at top ion-SPS energy, 
but show much less suppression than at the top RHIC energy in the $\pi^{\pm}$ 
channel above $p_{{}_T}>2\GeVpc$. For other particle channels, the behavior of the 
$R_{CP}^{BC}$ points at SPS and at RHIC seem to be rather similar, except for $\bar{p}$, 
which may be affected by the larger systematic errors due to the antiproton 
detection technique at the experiment NA49. The pQCD-based energy loss model seems to give 
a fair description of the $R_{CP}$ data points at SPS energy.

\begin{figure}[!ht]
\begin{center}
\includegraphics{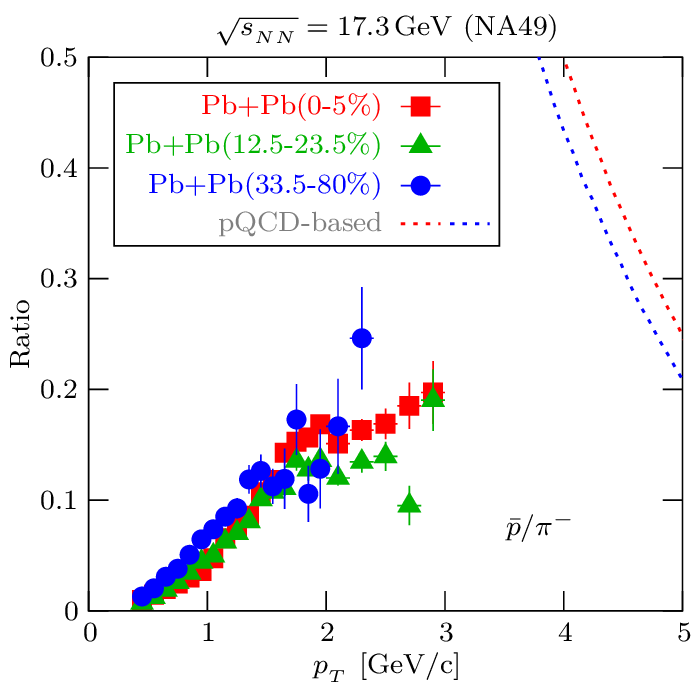}
\end{center}
\caption{Produced-baryon to meason ratios in Pb+Pb collisions, at 
$\sqrt{s_{{}_{NN}}}=17.3\,\mathrm{GeV}$ \cite{hipt}. The dotted lines 
indicate a pQCD-based energy loss model prediction \cite{xnw}.}
\label{BMR}
\end{figure}

To further test the predictions of the pQCD-based model, other particle ratios 
may also be looked at. It is found that the $\bar{p}/\pi^{-}$ data are not reproduced, 
as shown in Fig.\ \ref{BMR}.

\begin{figure}[!ht]
\begin{center}
\includegraphics{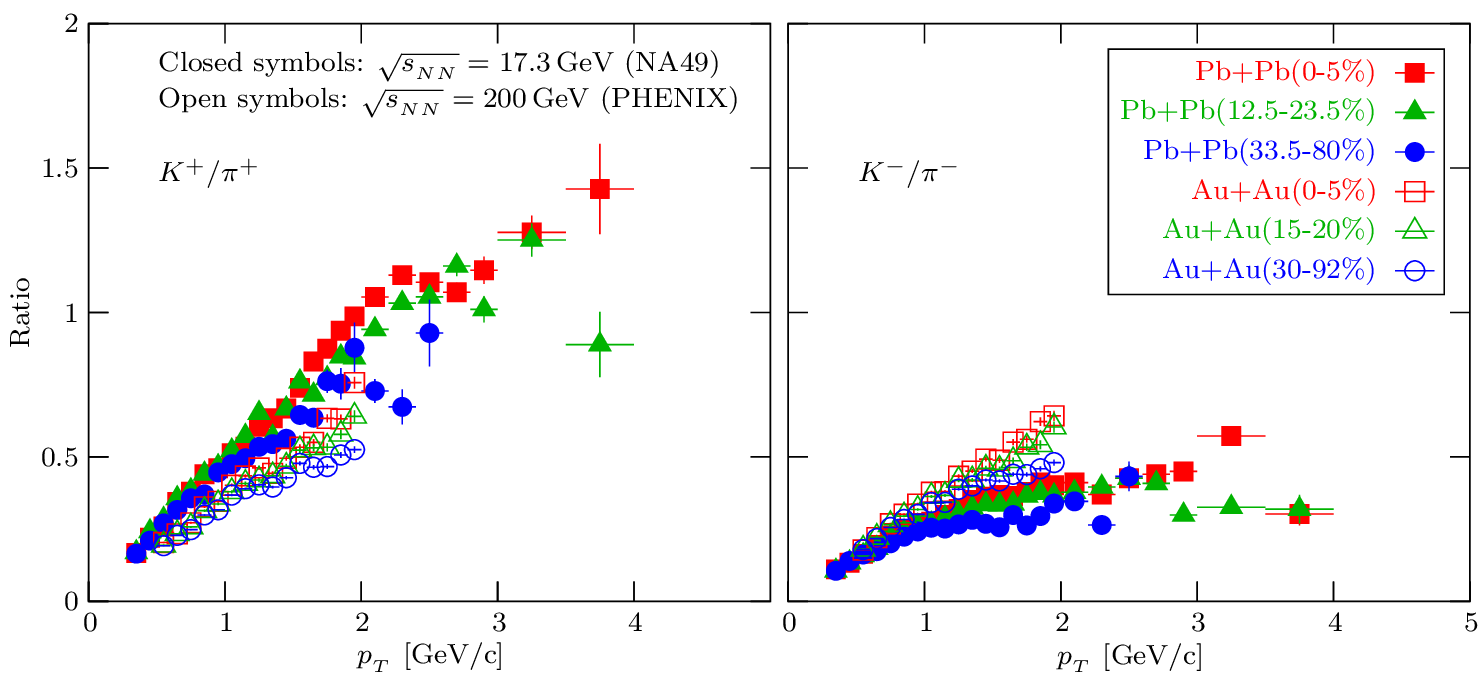}
\end{center}
\caption{
$K^{+}/\pi^{+}$ and $K^{-}/\pi^{-}$ ratios in Pb+Pb collisions, at 
$\sqrt{s_{{}_{NN}}}=17.3\GeV$ \cite{hipt} and at 
$\sqrt{s_{{}_{NN}}}=200\GeV$ \cite{phenixauaupid}.}
\label{SR}
\end{figure}

To get information about the strangeness production differentiated in 
centrality and $p_{{}_T}$, the $K^{+}/\pi^{+}$ and $K^{-}/\pi^{-}$ 
ratios were also extracted from the data, and are compared to 
$\sqrt{s_{{}_{NN}}}=200\GeV$ data \cite{phenixauaupid}. All 
this is shown in Fig.\ \ref{SR}. A monotonic increase of both ratios 
with centrality and $p_{{}_T}$ is observed. The $K^{-}/\pi^{-}$ points 
at top ion-SPS energy are slightly below the top RHIC energy points, whereas 
the $K^{+}/\pi^{+}$ points at top ion-SPS energy are above the RHIC points.

\section{Summary}

The measured $R_{AA}$ and $R_{pA}$ data show a monotonic increase in the 
covered $p_{{}_T}$ region at SPS energy. The $R_{AA}^{BC}$ points stay below 
$R_{pA}^{BC}$. The $R_{AA/pA}^{BC}$ and $R_{CP}^{BC}$ stay well below unity 
at SPS energy for $\pi^{\pm}$, but much less suppression is observed than at RHIC 
above $p_{{}_T}>2\GeVpc$.

The $R^{W}$ data start from one and $R_{AA}^{W}$ is above $R_{pA}^{W}$ both 
at SPS and at RHIC energy.

The $R_{CP}^{BC}$ data follow the RHIC points closely, 
except for $\bar{p}$ particles and for $\pi^{\pm}$ at $p_{{}_T}>2\GeVpc$.

The $R_{CP}^{BC}$ points are explained by pQCD-based energy loss calculation 
as in \cite{xnw}, however the produced-baryon/meson ratios are not reproduced. 
Possibly pQCD is not applicable in this kinematic region.

The $K^{+}/\pi^{+}$ and $K^{-}/\pi^{-}$ ratios show a monotonic increase 
in centrality and $p_{{}_T}$ both at SPS and RHIC energy. The $K^{-}/\pi^{-}$ 
points are slightly below the RHIC points, whereas the $K^{+}/\pi^{+}$ points 
are above the points measured at RHIC energy.

\acknowledgments                                                                                  
                                                                                                  
This work was supported by                                                                        
the US Department of Energy Grant DE-FG03-97ER41020/A000,                                         
the Bundesministerium fur Bildung und Forschung, Germany,                                         
the Virtual Institute VI-146 of Helmholtz Gemeinschaft, Germany,                                  
the Polish Ministry of Science and Higher Education (1 P03B 006 30, 1 P03B 097 29, 1 P03B 121 29, 1 P03B 127 30), 
the Hungarian Scientific Research Fund (OTKA 68506),                                              
the Korea Research Foundation (KRF-2007-313-C00175),                                              
the Bulgarian National Science Fund (Ph-09/05),                                                   
the Croatian Ministry of Science, Education and Sport (Project 098-0982887-2878)                  
and Stichting FOM, the Netherlands.


\begin{thebibliography}{99}

  \bibitem{hipt} C.~Alt {\it et al.} (the NA49 Collaboration), {\it Phys.~Rev.}~{\bf C77} (2008) 034906.

  \bibitem{phenixauaunopid} S.~S.~Adler {\it et al.} (the PHENIX Collaboration), {\it Phys.~Rev.}~{\bf C69} (2004) 034909.

  \bibitem{phenixauaupid} S.~S.~Adler {\it et al.} (the PHENIX Collaboration), {\it Phys.~Rev.}~{\bf C69} (2004) 034910.

  \bibitem{phenixdaunopid} S.~S.~Adler {\it et al.} (the PHENIX Collaboration), {\it Phys.~Rev.~Lett.}~{\bf 91} (2003) 072303.

  \bibitem{phenixdaupid} S.~S.~Adler {\it et al.} (the PHENIX Collaboration), {\it Phys.~Rev.}~{\bf C74} (2006) 024904.

  \bibitem{starauaupid} L.~Ruan (the STAR Collaboration), {\it J.~Phys.}~{\bf G31} (2005) s1029.

  \bibitem{nim99} S.~Afanasiev {\it et al.} (the NA49 Collaboration), {\it Nucl.~Instr.~Meth.}~{\bf A430} (1999) 210.

  \bibitem{wa98} M.~M.~Aggarwal {\it et al.} (the WA98 Collaboration), {\it Eur.~Phys.~J.}~{\bf C23} (2002) 225.

  \bibitem{wa98new} M.~M.~Aggarwal {\it et al.} (the WA98 Collaboration), {\it Phys.~Rev.~Lett.}~{\bf 100} (2008) 242301.

  \bibitem{hipttech} A.~L\'aszl\'o, NA49 Technical Note (2007) [{\tt EDMS:879787}].

  \bibitem{pp} C.~Alt {\it et al.} (the NA49 Collaboration), {\it Eur.~Phys.~J.}~{\bf C45} (2006) 343.

  \bibitem{pC} C.~Alt {\it et al.} (the NA49 Collaboration), {\it Eur.~Phys.~J.}~{\bf C49} (2007) 897.

  \bibitem{pC2} C.~Alt {\it et al.} (the NA49 Collaboration), {\it Eur.~Phys.~J.}~{\bf C49} (2007) 919.

  \bibitem{VCALrel} A.~L\'aszl\'o, NA49 Technical Note (2006) [{\tt EDMS:815907}].

  \bibitem{VCALabs} A.~L\'aszl\'o, NA49 Technical Note (2007) [{\tt EDMS:885329}].

  \bibitem{cronin} D.~Antreasyan {\it et al.}, {\it Phys.~Rev.}~{\bf D19} (1979) 764.

  \bibitem{xnw} X.~N.~Wang, {\it Phys.~Lett.}~{\bf B595} (2004) 165.

\end{thebibliography}
\end{document}